\documentclass[epj]{svjour}
\usepackage{graphics}
\usepackage{epsfig}
\begin{document}
\title{Electroproduction of nucleon resonances}
\author{L. Tiator\inst{1}, D. Drechsel\inst{1}, S. Kamalov\inst{1,2},
M.M. Giannini\inst{3}, E. Santopinto\inst{3}, A. Vassallo\inst{3}
}                     % Do not remove

\authorrunning{L. Tiator at al.}%
\institute{Institut f\"ur Kernphysik, Johannes Gutenberg-Universit\"at, D-55099 Mainz, Germany
\and JINR Dubna, 141980 Moscow Region, Russia \and
Dipartimento di Fisica dell'Universita di Genova and I.N.F.N., Sezione di Genova, I-16164 Genova,
Italy}
\date{Received: date / Revised version: date}
\abstract{ The unitary isobar model MAID has been extended and used for a partial
wave analysis of pion photo- and electroproduction in the resonance region $W<2$
GeV. Older data from the world data base and more recent experimental results from
Mainz, Bates, Bonn and JLab for $Q^2$ up to 4.0 (GeV/c)$^2$ have been analyzed and
the $Q^2$ dependence of the helicity amplitudes have been extracted for a series of
four star resonances. We compare single-$Q^2$ analyses with a superglobal fit in a
new parametrization of Maid2003 together with predictions of the hypercentral
constituent quark model. As a result we find that the helicity amplitudes and
transition form factors of constituent quark models should be compared with the
analysis of bare resonances, where the pion cloud contributions have been
subtracted.
} %end of abstract
\maketitle

\section{Introduction}
\label{introduction}

Our knowledge of nucleon resonances is mostly given by elastic pion nucleon
scattering \cite{Hohler79}. All resonances that are given in the Particle Data
Tables \cite{PDG02} have been identified in partial wave analyses of $\pi N$
scattering both with Breit-Wigner analyses and with speed-plot techniques. From
these analyses we know very well the masses, widths and the branching ratios into
the $\pi N$ and $\pi\pi N$ channels. These are reliable parameters for all
resonances in the 3 and 4 star categories. There remain some doubts for the two
prominent resonances, the Roper $P_{11}(1440)$ which appears unusually broad and the
$S_{11}(1535)$ that cannot uniquely be determined in the speed-plot due to its
position close to the $\eta N$ threshold. Both resonances, however, have recently
been found on the lattice in a very precise calculation with a pion mass as low as
$180$~MeV and converge very close to the empirical masses \cite{Frank03}. This has
been achieved in a quenched calculation giving rise to the conclusion that both
resonance states are simply $qqq$ states.

Starting from these firm grounds, using pion photo- and electroproduction we can
determine the electromagnetic $\gamma N N^*$ couplings. They can be given in terms
of electric, magnetic and charge transition form factors $G_E^*(Q^2)$, $G_M^*(Q^2)$
and $G_C^*(Q^2)$ or by linear combinations as helicity amplitudes $A_{1/2}(Q^2)$,
$A_{3/2}(Q^2)$ and $S_{1/2}(Q^2)$. So far, we have some reasonable knowledge of the
$A_{1/2}$ and $A_{3/2}$ amplitudes at $Q^2=0$, which are tabulated in the Particle
Data Tables. For finite $Q^2$ the information found in the literature is very scarce
and practically does not exist at all for the longitudinal amplitudes $S_{1/2}$. But
even for the transverse amplitudes only few results are firm, these are the $G_M^*$
form factor of the $\Delta(1232)$ up to $Q^2\approx 10$~GeV$^2$, the $A_{1/2}(Q^2)$
of the $S_{11}(1535)$ resonance up to $Q^2\approx 5$~GeV$^2$ and the asymmetry
$A(Q^2)=(A_{1/2}^2-A_{3/2}^2)/(A_{1/2}^2+A_{3/2}^2)$ for the $D_{13}(1520)$ and
$F_{15}(1680)$ resonance excitation up to $Q^2\approx 3$~GeV$^2$ which change
rapidly between $-1$ and $+1$ at small $Q^2\approx 0.5$~GeV$^2$ \cite{Boffi96}.
Frequently also data points for other resonance amplitudes, e.g. for the Roper are
shown together with quark model calculations but they are not very reliable. Their
statistical errors are often quite large but in most cases the model dependence is
as large as the absolute value of the data points. In this context it is worth noting that also the
word `data point` is somewhat misleading because these photon couplings and amplitudes cannot
be measured directly but can only be derived in a partial wave analysis. Only in the
case of the $\Delta(1232)$ resonance this can and has been done directly in the
experiment by Beck et al. at Mainz \cite{Beck00}. For the Delta it becomes possible
due to two important theoretical facts, the Watson theorem and the well confirmed
validity of the $s$+$p$ -- wave truncation. Within this assumption a complete
experiment was done with polarized photons and with the measurement of both $\pi^0$
and $\pi^+$ in the final state, allowing also for an isospin separation. For
other resonances neither the theoretical constraints are still valid nor are we any
close to a complete experiment. The old data base was rather limited with large
error bars and no data with either target or recoil polarization was available. Even
now we do not have many data points with double polarization, however, the situation
for unpolarized $e+p\rightarrow e'+p+\pi^0$ has considerably improved, mainly by the new JLab
experiments in all three halls A,B and C. These data cover a large energy range from
the Delta up to the third resonance region with a wide angular range in
$\theta_\pi$. Due to the $2\pi$ coverage in the $\phi$ angle a separation of the
unpolarized cross section
\begin{equation}
\frac{d\sigma_v}{d\Omega}= \frac{d\sigma_T}{d\Omega} + \varepsilon\frac{d\sigma_L}{d\Omega}
+\sqrt{2\varepsilon(1+\varepsilon)}\frac{d\sigma_{LT}}{d\Omega}\cos\phi
+\varepsilon\frac{d\sigma_{TT}}{d\Omega}\cos 2\phi
\end{equation}
in three parts becomes possible and is very helpful for the partial wave analysis.
Even without a Rosenbluth separation of $d\sigma_T$ and $d\sigma_L$ we have an
enhanced sensitivity of the longitudinal amplitudes due to the $d\sigma_{LT}$
interference term. Such data are the basis of our new partial wave analysis with an
improved version of the Mainz unitary isobar model MAID.

\section{Photo- and electroproduction}
\label{Photo- and electroproduction}

For our analysis of pion electroproduction we will use the dynamical model DMT
\cite{DMT} and the unitary isobar model MAID \cite{Maid}. In the dynamical approach
to pion photo- and electroproduction \cite{Yang85}, the t-matrix is expressed as
\begin{eqnarray}
t_{\gamma\pi}(E)=v_{\gamma\pi}+v_{\gamma\pi}\,g_0(E)\,t_{\pi N}(E)\,, \label{eq:tgamapi}
\end{eqnarray}
where $v_{\gamma\pi}$ is the transition potential operator for $\gamma^*N \rightarrow \pi N$, and
$t_{\pi N}$ and $g_0$ denote the $\pi N$ t-matrix and free propagator, respectively, with $E \equiv
W$ the total energy in the CM frame. A multipole decomposition of Eq. (\ref{eq:tgamapi}) gives the
physical amplitude in channel $\alpha$~\cite{Yang85},
\begin{eqnarray}
&&t_{\gamma\pi}^{(\alpha)}(q_E,k;E+i\epsilon) =\exp{(i\delta^{(\alpha)})}\,\cos{\delta^{(\alpha)}}
\times [v_{\gamma\pi}^{(\alpha)}(q_E,k)\nonumber\\ && \qquad+P\int_0^{\infty} dq' \frac{q'^2R_{\pi
N}^{(\alpha)}(q_E,q';E)\,v_{\gamma\pi}^{(\alpha)}(q',k)}{E-E_{\pi N}(q')}]\,, \label{eq:backgr}
\end{eqnarray}
where $\delta^{(\alpha)}$ and $R_{\pi N}^{(\alpha)}$ are the $\pi N$ scattering phase shift and
reaction matrix in channel $\alpha$, respectively; $q_E$ is the pion on-shell momentum and $k=|{\bf
k}|$ is the photon momentum. The multipole amplitude in Eq. (\ref{eq:backgr}) manifestly satisfies
the Watson theorem and shows that the $\gamma\pi$ multipoles depend on the half-off-shell behavior
of the $\pi N$ interaction.

In a resonant channel the transition potential $v_{\gamma\pi}$ consists
of two terms
\begin{eqnarray}
v_{\gamma\pi}(E)=v_{\gamma\pi}^B +v_{\gamma\pi}^R(E),\label{eq:vgammapi33}
\end{eqnarray}
where $v_{\gamma\pi}^B$ is the background transition potential and
$v_{\gamma\pi}^R(E)$ corresponds to the contribution of the
bare resonance excitation. The resulting t-matrix can be decomposed
into two terms \cite{KY99}
\begin{eqnarray}
t_{\gamma\pi}(E)=t_{\gamma\pi}^B(E) +
t_{\gamma\pi}^{R}(E),\label{eq:tgammapi33}
\end{eqnarray}
where
\begin{eqnarray}
t_{\gamma\pi}^B(E)=v_{\gamma\pi}^B+v_{\gamma\pi}^B\,g_0(E)\,t_{\pi
N}(E), \\
t_{\gamma\pi}^R(E)=v_{\gamma\pi}^R+v_{\gamma\pi}^R\,g_0(E)
\,t_{\pi N}(E).
\end{eqnarray}
Here $t_{\gamma\pi}^B$ includes the contributions from the nonresonant background
and renormalization of the  vertex $\gamma^*NR$. The advantage of such a
decomposition is that all the processes which start with the excitation of a bare
resonance are summed up in $t_{\gamma\pi}^R$. Note that the multipole decomposition
of both $t_{\gamma\pi}^B$ and $t_{\gamma\pi}^R$ would take the same form as Eq.
(\ref{eq:backgr}).

As in MAID \cite{Maid}, the background  potential $v_{\gamma\pi}^{B,\alpha}(W,Q^2)$
was described by Born terms obtained with an energy dependent mixing of
pseudovector-pseudoscalar $\pi NN$ coupling and t-channel vector meson exchanges.
The mixing parameters and coupling constants were determined from an analysis of
nonresonant multipoles in the appropriate energy regions. In the new version of
MAID, the $S$, $P$, $D$ and $F$ waves of the background contributions are unitarized
in accordance with the K-matrix approximation,
\begin{equation}
 t_{\gamma\pi}^{B,\alpha}({\rm MAID})=
 \exp{(i\delta^{(\alpha)})}\,\cos{\delta^{(\alpha)}}
 v_{\gamma\pi}^{B,\alpha}(W,Q^2).
\label{eq:bg00}
\end{equation}

From Eqs. (\ref{eq:backgr}) and  (\ref{eq:bg00}), one finds that the difference
between the background terms of MAID and of the dynamical model is that off-shell
rescattering contributions (principal value integral) are not included in MAID. To
take account of the inelastic effects at the higher energies, we replace
$\exp{(i\delta^{(\alpha)})} \cos{\delta^{(\alpha)}} = \frac 12
(\exp{(2i\delta^{(\alpha)})} +1)$ in Eqs. (\ref{eq:backgr}) and (\ref{eq:bg00}) by
$\frac 12 (\eta_{\alpha}\exp{(2i\delta^{(\alpha)})} +1)$, where $\eta_{\alpha}$ is
the inelasticity. In our actual calculations, both the $\pi N$ phase shifts
$\delta^{(\alpha)}$ and inelasticity parameters $\eta_{\alpha}$ are taken from the
analysis of the GWU group \cite{Arn97}.

Following  Ref.~\cite{Maid},  we assume a Breit-Wigner form for the resonance
contribution ${\cal A}^{R}_{\alpha}(W,Q^2)$ to the total multipole amplitude,
\begin{equation}
{\cal A}_{\alpha}^R (W,Q^2)\,=\, \bar{\cal A}_{\alpha}^R (Q^2)\, \frac{f_{\gamma
R}(W)\Gamma_R\,M_R\,f_{\pi R}(W)}{M_R^2-W^2-iM_R\Gamma_R} \,e^{i\phi}, \label{eq:BW}
\end{equation}
where $f_{\pi R}$ is the usual Breit-Wigner factor describing the decay of a
resonance $R$ with total width $\Gamma_{R}(W)$ and physical mass $M_R$. The
expressions for $f_{\gamma R}, \, f_{\pi R}$ and $\Gamma_R$ are given in
Ref.~\cite{Maid}. The phase $\phi(W)$ in Eq. (\ref{eq:BW}) is introduced to adjust
the phase of the total multipole to  equal  the corresponding $\pi N$  phase shift
$\delta^{(\alpha)}$. While in the original version of MAID \cite{Maid} only the 7
most important nucleon resonances were included with mostly only transverse e.m.
couplings, in our new version all four star resonances below $W=2$~GeV are included.
These are $P_{33}(1232)$, $P_{11}(1440)$, $D_{13}(1520)$, $S_{11}(1535)$,
$P_{33}(1232)$, $S_{31}(1620)$, $S_{11}(1650)$,
$D_{15}(1675)$, $F_{15}(1680)$,
$D_{33}(1700)$, $P_{13}(1720)$, $F_{35}(1905)$, $P_{31}(1910)$ and $F_{37}(1950)$.

\begin{table}[htb]
\caption{Recent experimental data of $\pi^0$ electroproduction on the proton. The
Mainz experiment was done with beam and recoil polarization, all others are
unpolarized measurements. From JLab Hall B data sets at fixed $Q^2$ of 0.4, 0.525,
0.65, 0.75 0.90, 1.15 and 1.45 $GeV^2$ have been used.} \label{tab:3}
\begin{center}
\begin{tabular}{l|lll}
\hline  laboratory & $Q^2 (GeV^2)$ & $W_{cm} (MeV)$ & $\theta_\pi^{cm} (deg) $ \\ \hline Mainz
\cite{Pos01}& 0.121 &  1232       & $180^\circ$ \\ Bates \cite{Mer01}& 0.126 & 1152 - 1322 &  $0 -
38^\circ$ \\ Bonn \cite{Ban02}& 0.630 & 1153 - 1312 &  $ 5 - 175^\circ$ \\ JLab, Hall
A\cite{Lav01}& 1.0 & 1110 - 1950 &  $ 146 - 167^\circ$ \\ JLab, Hall B\cite{Joo02}& $0.4 - 1.8$ &
1100 - 1680 &  $ 26 - 154^\circ$ \\ JLab, Hall C\cite{Fro99}& $2.8, 4.0$ & 1115 - 1385 &  $ 25 -
155^\circ$ \\ \hline
\end{tabular}
\end{center}
\end{table}

The resonance couplings $\bar{\cal A}_{\alpha}^R(Q^2)$ are independent of the total
energy and depend only on $Q^2$. They can be taken as constants in a single-Q$^2$
analysis, e.g. in photoproduction, where $Q^2=0$ but also at any fixed $Q^2$, where
enough data with W and $\theta$ variation is available. Alternatively they can also
be parametrized as functions of $Q^2$ in an ansatz like
\begin{equation}
\bar{\cal A}_{\alpha}(Q^2) = \bar{\cal A}_{\alpha}(0) \frac{1+c_1^\alpha Q^2}{(1+c_2^\alpha Q^2)^n}
\end{equation}
with $n\geq 2$. Also other parameterizations with an asymptotic fall-off for large $Q^2$,
e.g. of Gaussian form work equally well and can effectively be rewritten in either form.
With such an ansatz it is possible to determine the parameters
$\bar{\cal A}_{\alpha}(0)$ from a fit to the world database of photoproduction,
while the parameters $c_1^\alpha$ and $c_2^\alpha$ can be obtained from a combined
fitting of all electroproduction data at different $Q^2$. The latter procedure we
call the `superglobal fit'. In MAID the photon couplings $\bar{\cal A}_{\alpha}$ are
direct input parameters.

Eq.(9) can also be used for a general definition. At the resonance position, $W=M_R$ we obtain
\begin{eqnarray}
{\cal A}_{\alpha}^R(M_R,Q^2)\, &=& i\,{\bar{\cal A}}_{\alpha}(Q^2) f_{\gamma R}(M_R) f_{\pi R}(M_R)
c_{\pi N} e^{i \phi(M_R)}\nonumber\\
&=& {\cal A}_{\alpha}^{res}(M_R,Q^2)\, e^{i \phi(M_R)}
\end{eqnarray}
with $f_{\gamma R}(M_R)=1$ and
\begin{equation}
f_{\pi R}(M_R)=\left[ \frac{1}{(2j+1)\pi} \frac{k_W}{|q|} \frac{m_N}{M_R} \frac{\Gamma_{\pi
N}}{\Gamma^2_{tot}} \right]^{1/2}\,.
\end{equation}
The factor $c_{\pi N}$ is $\sqrt{3/2}$ and $-1/\sqrt{3}$ for the isospin $3/2$ and isospin $1/2$
multipoles, respectively. This leads to the definition
\begin{equation}
\bar{\cal A}_{\alpha}(Q^2) = \frac{1}{c_{\pi N} f_{\pi R}(M_R)} \mbox{Im} {\cal A}_{\alpha}^{res}
(M_R, Q^2)\,.
\end{equation}
It is important to note that by this definition the phase factor $e^{i \phi}$ in
Eqs. (9 and 11) is not considered as part of the resonant amplitude but rather as an
artifact of the unitarization procedure. In the case of the $\Delta(1232)$ resonance
this phase vanishes at the resonance position due to Watson's theorem, however, for
all other resonances it is finite and in some extreme cases it can reach values of
about $60^0$. $\bar{\cal A}_{\alpha}$ is a short-hand notation for the electric,
magnetic and longitudinal multipole photon couplings of a given partial wave
$\alpha$. As an example, for the $P_{33}$ partial wave the specific couplings are
denoted by $\bar{E}_{1+}$, $\bar{M}_{1+}$ and $\bar{S}_{1+}$. By linear combinations
they are connected with the more commonly used helicity photon couplings $A_{1/2}$,
$A_{3/2}$ and $S_{1/2}$.

For resonances with total spin $j=\ell + 1/2$ we get
\begin{eqnarray}
A^{\ell +}_{1/2} &=& -\frac{1}{2} [(\ell +2) \bar{E}_{\ell+} + \ell \bar{M}_{\ell+} ]\,, \nonumber\\
A^{\ell +}_{3/2} &=& \frac{1}{2}\sqrt{\ell(\ell+2)} (\bar{E}_{\ell+} - \bar{M}_{\ell+})\,, \nonumber\\
S^{\ell +}_{1/2} &=& -\frac{\ell+1}{\sqrt{2}} \bar{S}_{\ell+}
\end{eqnarray}
and for $j=(\ell+1) - 1/2$
\begin{eqnarray}
A^{(\ell+1)-}_{1/2} &=& \frac{1}{2} [(\ell +2) \bar{M}_{(\ell+1)-} - \ell \bar{E}_{(\ell+1)-} ]\,, \nonumber\\
A^{(\ell+1)-}_{3/2} &=& -\frac{1}{2}\sqrt{\ell(\ell+2)} (\bar{E}_{(\ell+1)-} + \bar{M}_{(\ell+1)-})\,, \nonumber\\
S^{(\ell+1)-}_{1/2} &=& -\frac{\ell+1}{\sqrt{2}} \bar{S}_{(\ell+1)-}\,.
\end{eqnarray}

With some care, as will be discussed below, the photon couplings can be directly compared to matrix elements
of the electromagnetic current calculated in quark models between the nucleon and the excited
resonance states,
\begin{eqnarray}
A_{1/2} &=& -\sqrt{\frac{2\pi\alpha_{fs}}{k_W}}
<R,\frac{1}{2}\,|\,J_{+}\,|\,N,-\frac{1}{2}>\, \zeta\,, \nonumber\\
A_{3/2} &=& -\sqrt{\frac{2\pi\alpha_{fs}}{k_W}}
<R,\frac{3}{2}\,|\,J_{+}\,|\,N,\frac{1}{2}>\,\zeta\,, \nonumber\\
S_{1/2} &=& -\sqrt{\frac{2\pi\alpha_{fs}}{k_W}}
<R,\frac{1}{2}\,|\,\rho\,|\,N,\frac{1}{2}>\, \zeta \,,
\end{eqnarray}
where $J_{+}=-\frac{1}{\sqrt{2}}(J_x+iJ_y)\,$.
However, these couplings are only defined up to a phase
$\zeta$. Since the sign of the pionic decay of the resonance has been ignored in the
empirical definition of these amplitudes, Eq.~(13), it must be taken into account in
a model calculation in order to make comparison with the empirical data. Therefore
the phases $\zeta$ have to be individually calculated in each model. In the
calculations found in the literature this has often been ignored and causes some
confusion in comparing these numbers, especially in critical cases as for the Roper
resonance $P_{11}(1440)$, where the correct sign cannot simply be guessed.

\section{The hypercentral constituent quark model}
\label{sec:hcqm}

As an application for evaluating the photon couplings we have used the hypercentral
Constituent Quark Model (hCQM)~\cite{hqcm95}. It consists of a hypercentral quark interaction
containing a linear plus coulomb-like term, as suggested by lattice QCD calculations
\cite{bali}
\begin{equation}\label{eq:pot}
V(x)= -\frac{\tau}{x}~+~\alpha x~~,
 ~~\mathrm{with} ~~
~x=\sqrt{\mbox{\boldmath{$\rho$}}^2+{\mbox{\boldmath{$\lambda$}}}^2} ~~,
\end{equation}
%\noindent
where $x$ is the hyperradius defined in terms of the standard Jacobi coordinates
$\mbox{\boldmath{$\rho$}}$ and $\mbox{\boldmath{$\lambda$}}$. We can think of this
potential both as a two-body potential in the hypercentral approximation or as a
true three-body potential. A hyperfine term of the standard form \cite{is} is added
and treated as a perturbation. The parameters $\alpha$, $\tau$ and the strength of
the hyperfine interaction are fitted to the spectrum ($\alpha=1.61~fm^{-2}$,
$\tau=4.59$ and the strength of the hyperfine interaction is determined by the
$\Delta$ - Nucleon mass difference). Recently, isospin dependent terms have been
introduced~\cite{vass} in the hCQM hamiltonian. The complete interaction used is
given by
\begin{equation}\label{tot} H_{int}~=~V(x) +H_{\mathrm{S}} + H_{\mathrm{I}}
+H_{\mathrm{SI}}~.
\end{equation}
%where V(x) is the linear plus hypercoulomb SU(6)-invariant potential, already
%described in Eq.(17), while $H_{\mathrm{S}} + H_{\mathrm{I}} +H_{\mathrm{SI}}$, is a
%residual SU(6)-breaking interaction that can be treated as a perturbative term
%leading to an improved description of the spectrum.
Having fixed the values of all parameters, the resulting wave functions have been
used for the calculation of the photocouplings \cite{aie}, the transition form
factors for the negative parity resonances \cite{aie2}, the elastic form factors
\cite{mds,rap} and now also for the longitudinal and tranverse transition form
factors for all the 3- and 4-star and the missing resonances.
\begin{figure}[htb]
\centerline{\epsfig{file=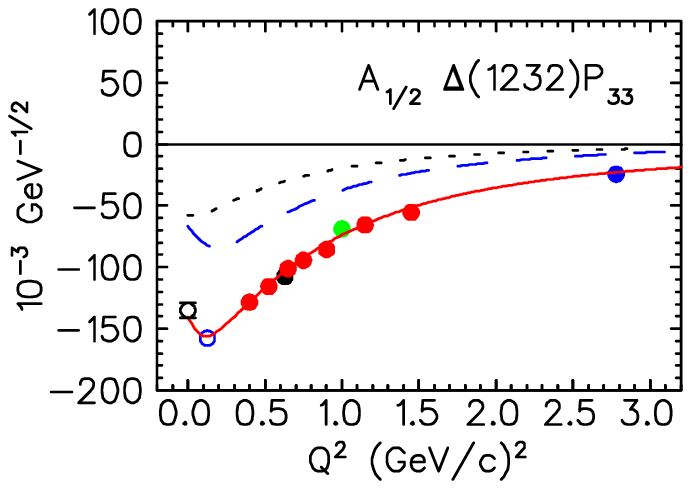,width=6.0cm,angle=0}} \vspace{0.3cm}
\centerline{\epsfig{file=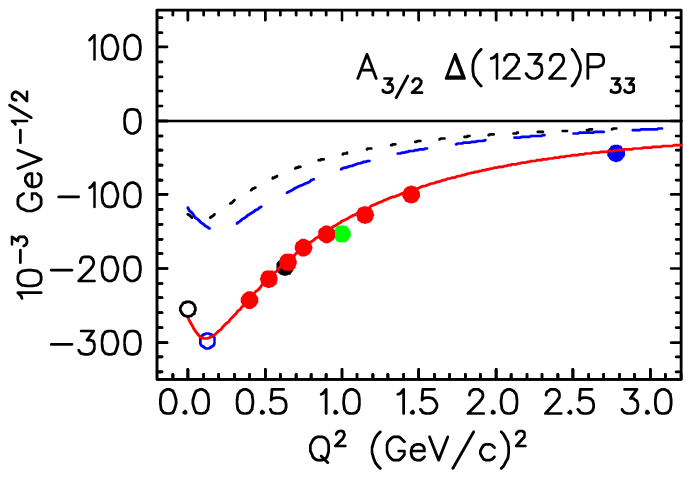,width=6.0cm,angle=0}} \vspace{0.3cm}
\centerline{\epsfig{file=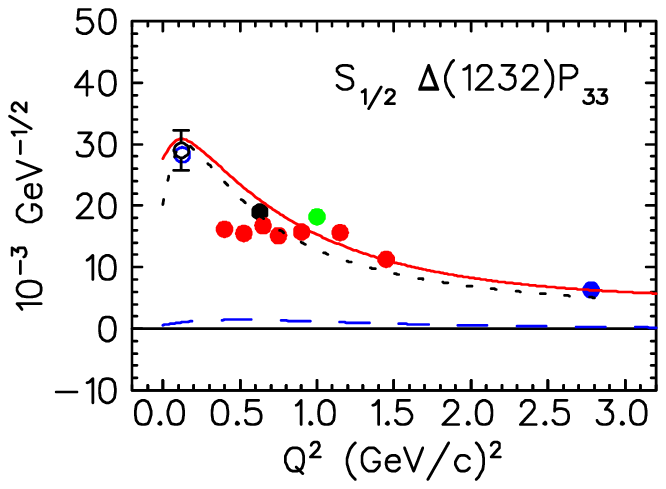,width=6.0cm,angle=0}} \caption{ The $Q^2$ dependence
of the $N \rightarrow \Delta$ helicity amplitudes. The solid and dashed curves are
the results of the superglobal fit with MAID and the predictions of the
hyperspherical constituent quark model. The dotted lines show the pion cloud
contributions calculated with DMT. The data points at finite $Q^2$ are the results
of our single-Q$^2$ fits, see Table 1 for references. At $Q^2=0$ for $A_{1/2}$ and
$A_{3/2}$ the photon couplings from PDG are shown \cite{PDG02}. }
\end{figure}
\label{fig:p33}

\section{Data analysis}
\label{sec:data analysis}

The unitary isobar model MAID was used to analyze the world data base of pion
photoproduction and recent differential cross section data on $p(e,e'p)\pi^0$ from
Mainz, Bates, Bonn and JLab. These data cover a $Q^2$ range from $0.1\cdots 4.0$
(GeV/c)$^2$ and an energy range $1.1 < W < 2.0$ GeV, see Table 1.  In a first
attempt we have fitted each data set at a constant $Q^2$ value separately. This is
similar to a partial wave analysis of pion photoproduction and only requires
additional longitudinal couplings for all the resonances. The $Q^2$ evolution of the
background, Born terms and vector meson exchange, is described with a standard
dipole form factor. In a second attempt we have introduced a $Q^2$ evolution of the
transition form factors of the nucleon to $N^*$ and $\Delta$ resonances and have
parameterized each of the transverse ($A_{1/2}$ and $A_{3/2}$) and longitudinal
($S_{1/2}$) helicity amplitudes. In a combined fit with all electroproduction data
from the world data base of GWU/SAID \cite{SAID} and the data of our single-$Q^2$
fit we obtained a $Q^2$ dependent solution (superglobal fit). In Fig. 1 we show our
results for the $\Delta(1232)$ excitation. Our superglobal fit agrees very well with
our single-$Q^2$ fits, except for the 2 lowest points of $S_{1/2}$ from our analysis
of the Hall B data. Whether this is an indication for a different $Q^2$ dependence
has still to be investigated. Generally, all our single-$Q^2$ points are shown with
statistical errors from $\chi^2$ minimization only. A much bigger error has to be
considered for model dependence.

We also compare our empirical analyses with the predictions of the hypercentral
constituent quark. It turns out that the transverse amplitudes of the quark model
are about half of the magnitudes and for the longitudinal amplitude $S_{1/2}(Q^2)$
the quark model gives essentially zero. This is due to the fact that in the
empirical analysis the resonance contribution is fully dressed by the pion cloud
which is not the case in a constituent quark model. As depicted in Fig. 2 a fully
dressed resonance contribution is renormalized on each vertex and in the propagator.
\begin{figure}[htb]
\centerline{\epsfig{file=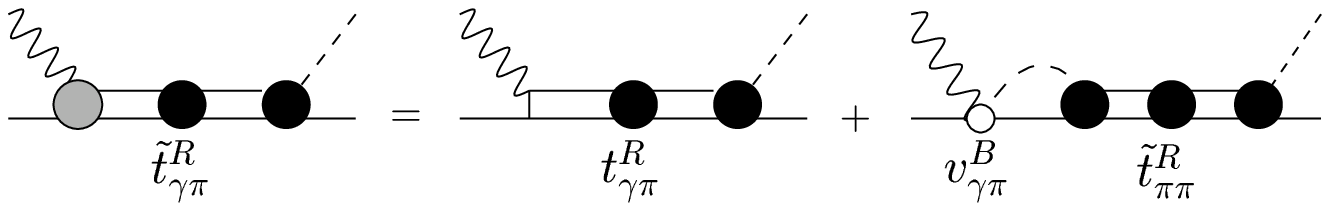,width=8cm,angle=0}}
\caption{Resonances with dressed and bare electromagnetic vertices.}
\end{figure}
\label{fig:dressed} The baryonic states of the hCQM including hyperfine interaction
can be considered as resonances dressed by hadronic interaction giving rise to the
empirical masses. However, the electromagnetic vertex correction (third part of
Fig.~2) is not included and has to be calculated separately. We have already started
to do this and in a first attempt we have used the dynamical model DMT and extracted
the pion loop contributions for the s- and p-waves. This estimate of the pion cloud
vertex correction is shown as dotted lines in the figures. For the longitudinal
Delta excitation the entire amplitude is practically given by the pion cloud
contribution and only a negligible part arises from a bare Delta. The same is true
for the electric amplitude which is given by the combination
$A_{1/2}-A_{3/2}/\sqrt{3}$. This calculation also explains why none of the
constituent quark models was ever able to give the empirical strength of the M1
excitation (or the transition moment $\mu_{N\Delta}$) of the $\Delta(1232)$. While
in a simple $SU(2)$ calculation the transition moment is lower by about $30\%$ in
more refined calculations it can be as low as only half of the empirical value. This
is also the case here in our hCQM, even if it has more realistic wave functions.
\begin{figure*}[ht]
\centerline{\epsfig{file=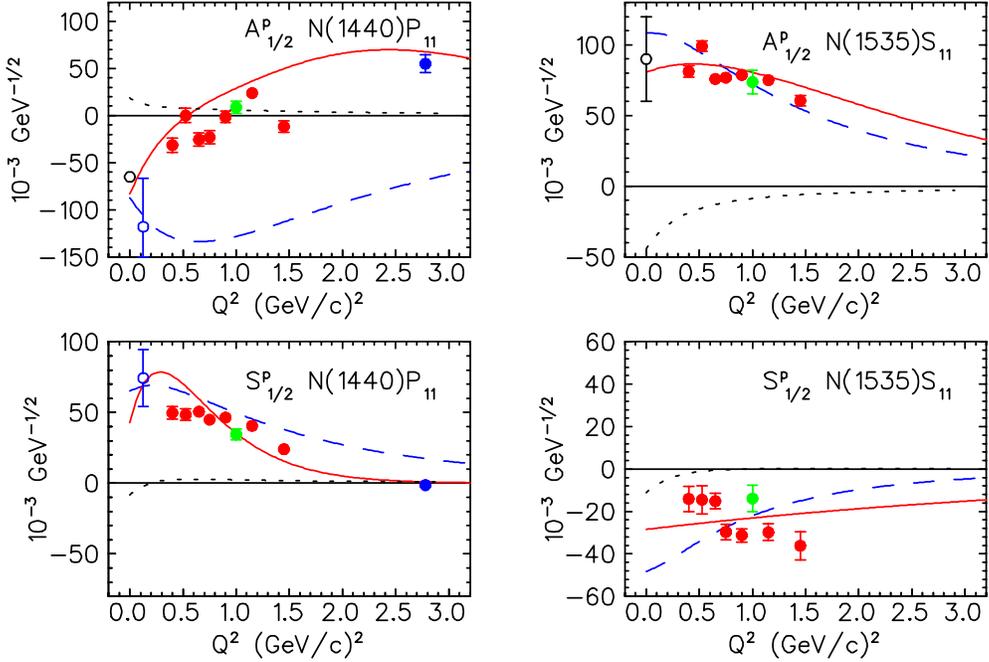,width=13cm,angle=0}}
\caption{ The $Q^2$ dependence of the transverse and longitudinal helicity
amplitudes for the $P_{11}(1440)$ and the $S_{11}(1535)$ resonance excitation.
The notation of the curves and the data is the same as in Fig. 1.}
\end{figure*}
\label{fig:s11p11}

In Fig. 3 we show our results for the helicity amplitudes of the Roper resonance
$P_{11}(1440)$ and the $S_{11}(1535)$. For these resonance amplitudes the pion cloud
contributions are most important near the photon point and become already negligible
around $Q^2=0.5$~GeV$^2$. The comparison between the hCQM and the empirical amplitudes
is reasonably good, except for the $A_{1/2}$ amplitude of the Roper. This finding
has to be further investigated both in the framework of the quark model and also in
the empirical analysis. Certainly, for the Roper resonance the existing data is not
very sensitive to this partial wave. Further experiments with double polarization
could be very helpful to solve this problem.
\begin{figure}[htb]
\centerline{\epsfig{file=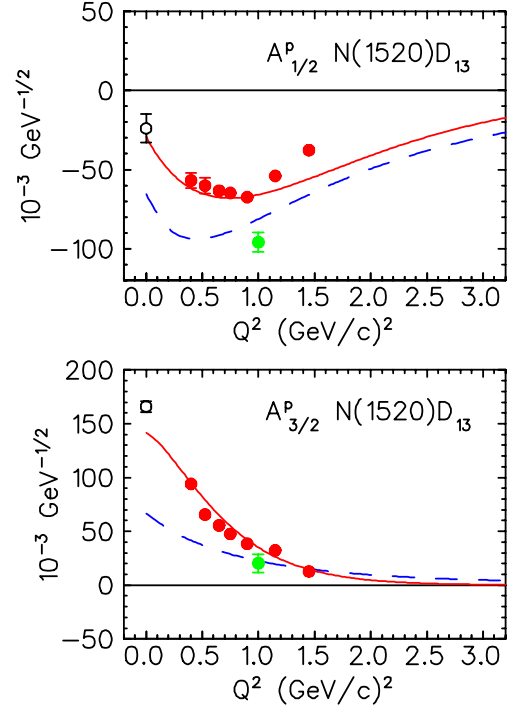,width=6.5cm,angle=0}} \caption{ The $Q^2$
dependence of the transverse helicity amplitudes for the $D_{13}(1520)$ resonance
excitation. The notation of the curves and the data is the same as in Fig. 1.}
\end{figure}
\label{fig:d13}

Finally, in Figs. 4 and 5 we show our results for the $D_{13}(1520)$ and the
$F_{15}(1680)$ resonances. Here the largest discrepancies between our quark model
calculations and the empirical analysis appear in the helicity $3/2$ amplitudes at
small $Q^2$. So far, the dynamical model calculations have only been done for $s$-
and $p$- waves, therefore we cannot give a pion cloud calculations for these partial
waves. However, our findings encourages very strongly such extensions of the
dynamical model. Furthermore we also have some empirical results for the partial
waves that are not shown here, but most of them come out of the fit with rather
large errors bars in the single-$Q^2$ analysis. This gives us less confidence also
for our superglobal fit. The reason for it is mainly that we have fewer data points to
analyze at higher energies.

\section{Conclusions}
\label{sec:concl}

Using the world data base of pion photo- and electroproduction and recent data from
Mainz, Bonn, Bates and JLab we have made a first attempt to extract all longitudinal
and transverse helicity amplitudes of nucleon resonance excitation for four star
resonances below $W=2$~GeV. For this purpose we have extended our unitary isobar
model MAID and have parametrized the $Q^2$ dependence of the transition amplitudes.
Comparisons between single-$Q^2$ fits and a $Q^2$ dependent superglobal fit give us
confidence in the determination of the Delta amplitudes. We can also reasonably well
determine the amplitudes of the $P_{11}(1440), S_{11}(1535), D_{13}(1520)$ and the
$F_{15}(1680)$, even though the model uncertainty of these amplitudes can be as
large as 50\% for the longitudinal amplitudes of the $D_{13}$ and $F_{15}$.
\begin{figure}[htb]
\centerline{\epsfig{file=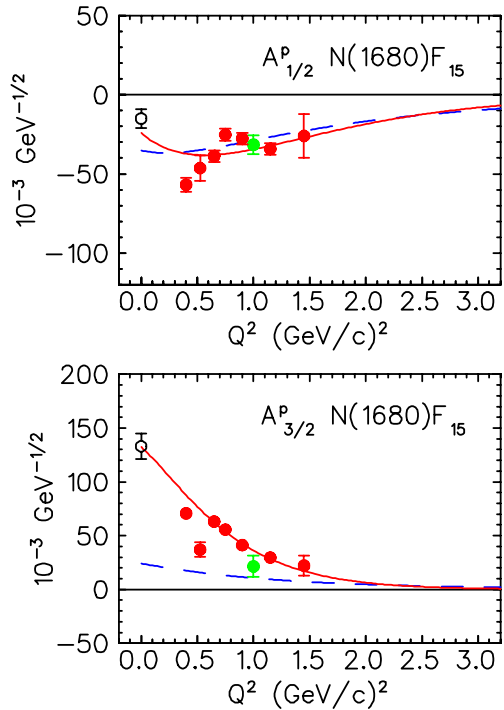,width=6.5cm,angle=0}}
\caption{ The $Q^2$ dependence of the transverse helicity
amplitudes for the $F_{15}(1680)$ resonance excitation.
The notation of the curves and the data is the same as in Fig. 1.}
\end{figure}
\label{fig:f15}
For other resonances the situation is even worse. However, this only
reflects the fact that precise data in a large kinematical range are absolutely
necessary. In some cases double polarization experiments are very helpful as has
already been shown in pion photoproduction. Furthermore, without charged pion
electroproduction, some ambiguities between partial waves that differ only in
isospin as $S_{11}$ and $S_{31}$ cannot be resolved without additional assumptions.
Finally, all results discussed here are only for the proton target. We have also
started an analysis for the neutron, where much less data are available from the
world data base and no new data has been analyzed in recent years. Since we can very
well rely on isospin symmetry, only the electromagnetic couplings of the neutron
resonances with isospin $1/2$ have to be determined. We have found such a solution
for the neutron and will implement it in the next version of MAID (MAID2003). It
will be a challenge for the experiment to investigate also the neutron resonances in
the near future.

\section{Acknowledgements}
We wish to thank T. Bantes, G. Laveissiere and C. Smith for their contribution on
the experimental data. This work was supported in part by the Deutsche
Forschungsgemeinschaft (SFB443).

\end{document}